\documentclass[aps,pre,twocolumn,array,epsfig,eqsecnum]{revtex4}
\usepackage{amsmath}
\usepackage{amsfonts}
\usepackage[dvips]{graphicx}
\usepackage{amsmath,tabu}

\begin{document}   
\setlength{\parindent}{0pt}

\title{Generalized Stokes vector for three photon process}

\author{Chitra Shaji, Raseena Ismail, SVM Satyanarayana, Alok Sharan}
\address{
Department of Physics, Pondicherry University, Puducherry 605 014, India}
\date{\today}

\begin{abstract}
Stokes Muller formalism is important to understand the optical properties of materials
by measuring the change in the polarization state of light upon scattering. The
formalism can be extended to nonlinear scattering processes involving two and three photon
processes. In this work, we derive a triple Stokes vector analytically using operator
approach used in quantum theory of light. A three photon polarization state can be
described by Stokes vector that has sixteen components involving cubes of intensities.
The response of a material for the scattering of light in a three photon process is
described by 4 x 16 Muller matrix. Polarization in Polarization out (PIPO) experiments
can be carried out to determine the elements of Muller matrix.
For that we identify 16 independent points in Poincare sphere and construct triple
Stokes vector for each point. Four measurements to find the linear Stokes
vector of the scattered light for incident light in each of the sixteen
three photon states determine the Muller matrix of the sample.
\end{abstract}

\maketitle

\section{Introduction}
General interactions of light with matter results in emission/ scattering of the light leading to observation of the Optical Phenomena. Different components detailing the description of wave e.g frequency, amplitude, speed, phase, polarization would require different approaches to understand these interactions.  Changes in the state of polarization of incident light beam are sensitive to the interaction with material and hence is used to predominantly characterize its optical properties.
Polarization state of light can be represented using Jones vector and Stokes vector where polarization properties of a material is represented using Mueller matrix formalism. Jones matrix of dimension $2\times1$ can represent only pure polarized light whereas unpolarized light, partially polarized light, completely polarized light can be represented using $4\times1$ Stokes vector. Each element of Stokes vector has the dimension of intensity, a very convenient quantity to determine experimentally. Optical properties of material can be described using Mueller matrix of dimension $4\times4$ which represent the incident and emerging light described by Stokes vector. Mueller matrix formalism is a preferred method to study systems in which scattering, depolarization, diattenuation etc occurs. Mueller matrix of a material is wavelength specific and is dependent on the scattering geometry at which measurement is done. Here we present Stokes formalism which can be used to study three photon nonlinear process \cite{polarbook,collet84,Bickel85,Azzam85,Simon92,muellerseawater,NirmalyaJBO,handbookosa}.\newline

 Mueller matrices have been widely used in studying linear scattering process \cite{harris92,NirmalyaJBO,muellerseawater,angular}. Utility of Mueller matrix measurement has found wide applications in biophotonics, remote sensing, radar polarimetry, and understanding different optical systems \cite{applicationbook}. Given an optically active material, its depolarization, birefringence, diattenuation properties can be effectively determined using this technique \cite{decomposition96,applicationbook}. First attempt to use Stokes Mueller formalism to study polarization dependance of nonlinear scattering is by Y. Shi et.al \cite{yshiscatter94,Yshi1994}. When light of high enough intensity falls on a material, its optical interaction could be nonlinear. Depending on the incident intensity, symmetry of the material and the orientation of the crystal axis with respect to the propagation direction of light, the material will of nonlinear optical processes of different order\cite{boyd}. The first formalism described by Y.Shi et.al, \cite{yshiscatter94} for two photon nonlinear process, derives two photons from two different beam and had constructed a $4\times4\times4 $ super Mueller matrix to explain the nonlinear scattering in a medium in which the $\hat{k}$ of incident photons and the emitted one are coplanar. In \cite{Yshi1994}, the author has considered the nonlinear scattering in which two photons are taken simultaneously from a single beam. A $9\times1$ Stokes vector along with $4\times9$ Mueller matrix explains  nonlinear scattering such as two photon absorption, frequency doubling, hyper Raman scattering etc. Second harmonic generation microscopy using $4\times1$ Stokes vector and $4\times4$ Mueller matrix has been done in biological materials to reveal structures of the materials in molecular level \cite{stokenirmalshg,shgnirmal,revealshg,stokesshg}. The properties of a SHG (Second Harmonic Generation) from biological structures has been recently studied using  double Stokes vector and Double Mueller matrix by M. Samim et.al, Lukas Kontenis et.al, \cite{Masood2015,Msamimconf,samimconf2}. Double Stokes Mueller Polarimetry experiments done in collagen of rat tail tendon is effective in cancer diagnosis \cite{Msamimconf}. A relation between the susceptibility components and the double Mueller matrix elements have been established by M.Samim {\it{et.al.,}} \cite{Masood2015}. Recently M. Samim \textit{et.al.,} has given a generalized form Stokes-Mueller which explains nonlinear process such as second harmonic generation, third harmonic generation, sum-difference frequency generation based on the classical description of electric field \cite{masoodarchive}. Where as we have used the quantum mechanical approach to derive the generalized form of triple Stokes vector which can explain all three photon process along with triple Mueller matrix. A generalized relation which  connects any order susceptibility with its corresponding Mueller matrix elements has been recently reported \cite{masoodarchive}.\newline
An effective formalism explaining three photon nonlinear process using Stokes-Mueller formalism is not found in literature. Here we present the $16\times1$ Stokes vector which can be used to study a three photon process. We consider the simultaneous anhilation creation of three photons from a single beam. Nonlinear process such as third harmonic generation, sum frequency generation, saturable absorption, coherent anti Stokes Raman scattering can be explained using $16\times1$ stokes vector along with $4\times16$ triple Mueller matrix. Since third order optical nonlinearity is  universal optical nonlinearity and since almost all materials show this nonlinearity, this formalism helps to determine the third order susceptibility components of these materials. The major advantage in using Mueller matrix formalism in nonlinear scattering is that, all susceptibility components present in given orientation of the crystal can be extracted simultaneously.\newline
 Section II of the paper describes the derivation of Stokes operators using quantized form of light \cite{quantumfluc,luis2000}. We have used same approach for obtaining linear Stokes vector, double Stokes vector and triple Stokes vector. The formulation and derivation of the sixteen triple Stokes vectors which have cubic power of incident intensity are presented in section III. Further the reduced density matrix of the Stokes vector is also obtained. This density matrix is significant when we consider physical process such as third harmonic generation, sum frequency generation etc. Also we describe a Polarization In Polarization Out (PIPO) arrangement for the measurement of the $4\times16$ Mueller matrix of the sample. We have shown how sixteen basis points are chosen from the Poincare sphere for experiment. These sixteen points lie symmetrically on the surface of Poincare sphere. The linear Stokes vector and triple Stokes vector of sixteen basis vectors are also given.
\section {Linear and Double Stokes vectors}
Optical process involving n simultaneous photons, $(n+1)^{2}$ independent polarization parameters are required to characterize the process \cite{Yshi1994}. 16 independent polarization parameters are required to study three photon process, which is the focus of our present work. Before getting into the formulation of three photon Stokes parameter, a brief review of formulation of linear Stokes operators and double Stokes operators is given. \newline
Light as an electromagnetic wave can be primarily described by its electric field in most of its interactions with materials  to explain the resulting optical phenomenon. Consider the quantized electric field of light. Any arbitrary electric field $E(r,t)$ can be represented as summation of positive and negative frequency parts. $E(r,t)=E^+(r,t)+ E^-(r,t)$ where + and - denotes the positive and negative frequencies, r is the propagation direction of light.
Let us consider the  monochromatic field with frequency $\omega$ propagating along z direction and the light is polarized along x - y plane. Using plane wave decomposition the positive frequency component of electric field can be written as
\begin{equation}
\vspace{0.2cm}
E^+(z,t)= (\frac{2\pi\hbar\omega}{n^2 (\omega)V})exp[-i(\omega t-kz)] \sum_{\lambda\approx 1,2} e^{\lambda}\hat{a}_{\lambda}
\vspace{0.2cm}
\end{equation}
$n(\omega)$ is the refractive index of the medium, V is the quantization volume, $k=\frac{n(\omega)\omega}{c}$\cite{quantumfluc}, $\hat{a}$ is the anhilation operator. The adjoint of $E^+(z,t)$ that is, $E^-(z,t)$, contains creation operator $a^+$. In case of cartesian basis, we have $\hat{e}^1=\hat{x}$, $\hat{e}^{2}=\hat{y}$.The two mode description of electric field helps to obtain quantum Stokes operators.The anhilation, creation operators $({a}_{x}^{\pm})$ in the x direction creates or destroys a horizontally polarized photon, hence we can denote it as $a_{H}^{\pm}$. The same is applicable for vertically polarized photon and hence denoted as $a_{V}^{\pm}$.\newline


Let $a_H ^\pm$ and $a_V^\pm$ represent anhilation creation operators of horizontally as well as vertically polarized photons respectively. By using the Pauli matrices $(\tau_\alpha, \alpha=1,2,3,4)$ along with the identity matrix, $4\times1$ Stokes vector can be expressed in the operator form. The expectation value of the Stokes operator is found using $S=Tr(\rho\hat{S})$. The general form of linear Stokes vector can be defined as \cite{Yshi1994}.\newline
\begin{equation}
\hat{s}_{\alpha}=\sum_{i=H}^{V}\sum_{j=H}^{V}(\tau_{\alpha})_{ij}a_i^+a_j^-(\alpha=1,2,3,4)\\
\end{equation}
\begin{eqnarray}
\tau_1=\left(
         \begin{array}{cc}
           1 & 0 \\
           0 & 1 \\
         \end{array}
       \right)
\tau_2=\left(
         \begin{array}{cc}
           1 & 0 \\
           0 & -1 \\
         \end{array}
       \right)\\\nonumber
\tau_3=\left(
         \begin{array}{cc}
           0 & 1 \\
           1 & 0 \\
         \end{array}
       \right)
\tau_4=\left(
         \begin{array}{cc}
           0 & i \\
           -i & 0 \\
         \end{array}
       \right)
\end{eqnarray}
where it obeys orthogonality $[\tau_{\alpha}\tau_{\beta}]=2\delta_{\alpha\beta}$. The linear Stokes operator is given as following
\begin{eqnarray}
       \hat{s}_0 &=& \quad a_H^+a_H^-+a_V^+a_V^-\nonumber\\
       \hat{s}_1 &=& \quad a_H^+a_H^--a_V^+a_V^-\nonumber\\
       \hat{s}_2 &=& \quad a_H^+a_V^--a_V^+a_H^-\nonumber\\
       \hat{s}_3 &=& \quad -i(a_H^+a_V^--a_V^+a_H^-)\\\nonumber
\end{eqnarray}
The expectation value of these Stokes operator with respect to the photon density matrix \cite{Yshi1994} gives the Stokes vector. The  double Stokes vector is derived analogous to the derivation of linear Stokes vector. Double creation and double anhilation operators are used here. These are defined as follows:
\begin{equation}
 b_1^{\pm}\equiv a_H^{\pm}a_H^{\pm}\qquad  b_2^{\pm}\equiv a_V^{\pm}a_V^{\pm}\qquad
        b_3^{\pm}\equiv \sqrt{2}a_{H}^{\pm}a_{V}^{\pm}
\end {equation}
Double Stokes vector is given in a general form in the following way
\begin{equation}
\hat{S}_{\zeta}=\sum_{\mu,\nu=1,1}^{3,3}(\lambda_{\zeta})_{\mu,\nu}b_{\mu}^{+}b_{\nu}^{-}
\end{equation}

where $\zeta$ runs from 1-9,  9 $\lambda$ matrices are Gell-Mann matrices. The double Stokes vector can be obtained \cite{Yshi1994} by taking expectation value of the double Stokes vector. The nonlinear optical process such as second harmonic generation, sum frequency generation, two photon absorption, etc involving two photons can be explained using above vector.\newline

\section{Derivation of Triple Stokes Vectors}

Now let us consider an optical process in which three photons are involved. We use a $16\times1$ Stokes vector along with $4\times16$ Mueller matrix in order to understand three photon process occurring in a material. We derive the three photon Stokes vector analogous to how double Stokes vector and linear Stokes vector is derived.Let us start by considering triple anhilation creation operators. In order to construct the $16$ Stokes vectors, we have used operators that create and annihilate three identical photons simultaneously. The  four different triple annihilation or creation operators used to generate the $16$ Stokes vector are given as following,
\begin{eqnarray}
C_{1}^{\pm} = a_{H}^\pm a_{H}^\pm a_{H}^\pm  \qquad  C_{2}^{\pm} = a_{V}^\pm a_{V}^\pm a_{V}^\pm\\\nonumber
\qquad C_{3}^{\pm} = \sqrt{3}a_{H}^\pm a_{V}^\pm a_{V}^\pm \quad  C_{4}^{\pm} = \sqrt{3} a_{H}^\pm a_{H}^\pm a_{V}^\pm\nonumber
\end{eqnarray}
where H and V are basis vectors which span through all other polarization states. The general Stokes vector can be written as the following.
\begin{equation}
\hat{S}_{N}=\sum_{\mu,\nu=1,1}^{4,4}(\Omega_{N})_{\mu,\nu}C_{\mu}^{+}C_{\nu}^{-}
\end{equation}
where N varies from 1-16 and $(\mu,\nu)$ goes from (1, 1) to (4, 4).The $\Omega$ matrices ($4\times4$) are Hermitian matrix generators of $SU(4)$ analogous to Pauli matrices $(SU(2))$ and Gell Mann Matrices $SU(3)$ \cite{su4}.
\begin{widetext}
\begin{eqnarray}
    \Omega_1=\frac{1}{\sqrt{2}}
\left(
\begin{array}{cccc}
1 & 0 & 0 & 0 \\
0 & 1 & 0 & 0 \\
0 & 0 & 1 & 0 \\
0 & 0 & 0 & 1 \\
\end{array}
\right)
\Omega_{2} = \frac{1}{\sqrt{6}}\left(
\begin{array}{cccc}
1 & 0 & 0 & 0  \\
0 & 1 & 0 & 0  \\
0 & 0 & 1 & 0  \\
0 & 0 & 0 & -3 \\
\end{array}
\right)
\nonumber
\Omega_{3} = \frac{1}{\sqrt{3}}
\left(
\begin{array}{cccc}
1 & 0 &  0 & 0  \\
0 & 1 &  0 & 0  \\
0 & 0 & -2 & 0  \\
0 & 0 &  0 & 0  \\
\end{array}
\right)
\Omega_{4} =
\left(
\begin{array}{cccc}
1 & 0 & 0 & 0 \\
0 & -1 & 0 & 0 \\
0 & 0 & 0 & 0 \\
0 & 0 & 0 & 0 \\
\end{array}
\right)\\
\nonumber
\Omega_{5} =
\left(
\begin{array}{cccc}
0 & 1 & 0 & 0 \\
1 & 0 & 0 & 0 \\
0 & 0 & 0 & 0 \\
0 & 0 & 0 & 0 \\
\end{array}
\right)\nonumber
\Omega_{6} =
\left(
\begin{array}{cccc}
0 & 0 & 0 & 0 \\
0 & 0 & 1 & 0 \\
0 & 1 & 0 & 0 \\
0 & 0 & 0 & 0 \\
\end{array}
\right)\nonumber
\Omega_{7} =
\left(
 \begin{array}{cccc}
 0 & 0 & 0 & 0 \\
 0 & 0 & 0 & 0 \\
 0 & 0 & 0 & 1 \\
 0 & 0 & 1 & 0 \\
 \end{array}
 \right)\nonumber
 \Omega_{8} =
 \left(
 \begin{array}{cccc}
 0 & 0 & 1 & 0 \\
 0 & 0 & 0 & 0 \\
 1 & 0 & 0 & 0 \\
 0 & 0 & 0 & 0 \\
 \end{array}
 \right)\nonumber\\
 \Omega_{9} = \left(
 \begin{array}{cccc}
 0 & 0 & 0 & 0 \\
 0 & 0 & 0 & 1 \\
 0 & 0 & 0 & 0 \\
 0 & 1 & 0 & 0 \\
 \end{array}
 \right)\nonumber
 \Omega_{10} =
 \left(
  \begin{array}{cccc}
  0 & 0 & 0 & 1 \\
  0 & 0 & 0 & 0 \\
  0 & 0 & 0 & 0 \\
  1 & 0 & 0 & 0 \\
  \end{array}
  \right)\nonumber
  \Omega_{11} =
  \left(
                                           \begin{array}{cccc}
                                             0 & 0 & 0 & 0 \\
                                             0 & 0 & 0 & 0 \\
                                             0 & 0 & 0 & -i \\
                                             0 & 0 & i & 0 \\
                                           \end{array}
                                         \right)\nonumber
                                         \Omega_{12} = \left(
                                           \begin{array}{cccc}
                                             0 & -i & 0 & 0 \\
                                             i & 0 & 0 & 0 \\
                                             0 & 0 & 0 & 0 \\
                                             0 & 0 & 0 & 0 \\
                                           \end{array}
                                         \right)\nonumber\\
                                         \Omega_{13}=\left(
                                           \begin{array}{cccc}
                                             0 & 0 & -i & 0 \\
                                             0 & 0 & 0 & 0 \\
                                             i & 0 & 0 & 0 \\
                                             0 & 0 & 0 & 0 \\
                                           \end{array}
                                         \right)\nonumber
                                         \Omega_{14}=\left(
                                           \begin{array}{cccc}
                                             0 & 0 & 0 & 0 \\
                                             0 & 0 & -i & 0 \\
                                             0 & i & 0 & 0 \\
                                             0 & 0 & 0 & 0 \\
                                           \end{array}
                                         \right)\nonumber
                                         \Omega_{15}=\left(
                                           \begin{array}{cccc}
                                             0 & 0 & 0 & 0 \\
                                             0 & 0 & 0 & -i \\
                                             0 & 0 & 0 & 0 \\
                                             0 & i & 0 & 0 \\
                                           \end{array}
                                         \right)\nonumber
                                         \Omega_{16}=\left(
                                           \begin{array}{cccc}
                                             0 & 0 & 0 & -i \\
                                             0 & 0 & 0 & 0 \\
                                             0 & 0 & 0 & 0 \\
                                             i & 0 & 0 & 0 \\
                                           \end{array}
                                         \right)\nonumber\\
\end{eqnarray}
\end{widetext}
$\Omega$ matrices are  orthogonal and the orthogonality condition is given below
 \begin{equation}
 Tr(\Omega_{m}\Omega{n})=2\delta{mn}
 \end{equation}
$\Omega$ matrices also obey the characteristic commutation and Jacobi identity relation \cite{su4}.$[[\Omega_l,\Omega_m]\Omega_n] + [[\Omega_m,\Omega_n]\Omega_l] + [[\Omega_n,\Omega_l]\Omega_m]=0$.
The general expression of triple Stokes vector for three photon process is derived by expanding equation (3.2) in terms of creation anhilation operators. By using the expression of linear Stokes vector in terms of dagger operators (2.4), commutation relations (3.5), and by doing analytical calculations, we write the triple Stokes vector in terms of linear Stokes vector. The commutation relation for linear Stokes vector are given below.

\begin{align}
\nonumber
[s_{2},s_{3}]=2is_{4} \quad [s_{3},s_{4}]=2is_{2}\\
[s_{4},s_{2}]=2is_{3} \quad [s_{1},s_{\alpha}]=0\\
\nonumber
\alpha=2,3,4\\
\nonumber
\end{align}

Along with the commutation relations, expressions used to derive triple Stokes vectors are
\begin{align}
\nonumber
\hat{s}_2^2\hat{s}_3-\hat{s}_3^2\hat{s}_2^2=4i\hat{s}_{2}\hat{s}_{4}-4\hat{s}_{3}\\
\nonumber
\hat{s}_{3}\hat{s}_{4}+\hat{s}_{4}\hat{s}_{3}=2i\hat{s}_{2}+2\hat{s}_{4}\hat{s}_{3}\\
\hat{s}_{2}\hat{s}_{4}+\hat{s}_{4}\hat{s}_{2}=2i\hat{s}_{3}+2\hat{s}_{2}\hat{s}_{4}\\
\nonumber
\hat{s}_4^2\hat{s}_2-\hat{s}_2^2\hat{s}_4^2=4i\hat{s}_{4}\hat{s}_{3}-4\hat{s}_{2}\\
\nonumber
\end{align}
The derived triple Stokes vector is given below (3.7). The first column has the third power, second column has quadratic power of linear Stokes vector and third column is linear part which contain only first power of linear Stokes vector. The triple Stokes vectors are cubic in incident intensity.
\begin{widetext}
\begin{equation}
\begin{matrix}
\begin{bmatrix}
\hat{S}_{1}\\
\hat{S}_{2}\\
\hat{S}_{3}\\
\hat{S}_{4}\\
\hat{S}_{5}\\
\hat{S}_{6}\\
\hat{S}_{7}\\
\hat{S}_{8}\\
\hat{S}_{9}\\
\hat{S}_{10}\\
\hat{S}_{11}\\
\hat{S}_{12}\\
\hat{S}_{13}\\
\hat{S}_{14}\\
\hat{S}_{15}\\
\hat{S}_{16}\\
\end{bmatrix}
\end{matrix} =\qquad
\begin{bmatrix}
\frac{1}{\sqrt{2}}\hat{s}_{1}^{3}\\
\frac{1}{2\sqrt{6}}(-\hat{s}_{1}^{3}+3\hat{s}_{1}\hat{s}_{2}(\hat{s}_{2}-\hat{s}_{1})+3\hat{s}_{2}^{3}) \\
-\frac{1}{4\sqrt{3}}(3\hat{s}_{2}^{3}+2\hat{s}_{1}^{3})+\frac{\sqrt{3}}{4}\hat{s}_{1}\hat{s}_{2}(2\hat{s}_{2}+\hat{s}_{1})\\
 \frac{1}{4}\hat{s}_{2}(\hat{s}_{2}^{2}+3\hat{s}_{1}^{2})\\
\frac{1}{4}(\hat{s}_{3}^{3}-3\hat{s}_{3}\hat{s}_{4}^{2}) \\
\frac{\sqrt{3}}{4} (\hat{s}_{1}^{2} + \hat{s}_{2}^{2})\hat{s}_{3} -\frac{\sqrt{3}}{2}\hat{s}_{1}\hat{s}_{3}\hat{s}_{2}\\
 \frac{3}{4}(\hat{s}_{1}^{2}\hat{s}_{3}-\hat{s}_{2}\hat{s}_{3}\hat{s}_{2})\\
 \frac{\sqrt{3}}{4}\hat{s}_{3}^{2}(\hat{s}_{1}+\hat{s}_{2})-\frac{\sqrt{3}}{4}(\hat{s}_{1}+\hat{s}_{2})\hat{s}_{4}^{2}\\
 \frac{\sqrt{3}}{4}\hat{s}_{3}^{2}(\hat{s}_{1}-\hat{s}_{2})+\frac{\sqrt{3}}{4}(\hat{s}_{2}-\hat{s}_{1})\hat{s}_{4}^{2}\\
 \frac{\sqrt{3}}{4}(\hat{s}_{1}^{2}+\hat{s}_{2}^{2})\hat{s}_{3}+\frac{\sqrt{3}}{2}\hat{s}_{1}\hat{s}_{3}\hat{s}_{2}\\
 \frac{3}{4}(\hat{s}_{1}^{2}\hat{s}_{4}-\hat{s}_{2}\hat{s}_{4}\hat{s}_{2})\\
 \frac{1}{4}\hat{s}_{4}(\hat{s}_{4}^{2}-3\hat{s}_{3}^{2})\\
 \frac{i\sqrt{3}}{2}(i(\hat{s}_{2}+\hat{s}_{1})\hat{s}_{4}\hat{s}_{3})\\
 \frac{\sqrt{3}}{4}(\hat{s}_{1}^{2}+\hat{s}_{2}^{2})\hat{s}_{4}-\frac{\sqrt{3}}{2}\hat{s}_{1}\hat{s}_{2}\hat{s}_{4}\\
 \frac{i\sqrt{3}}{2}(i(\hat{s}_{2}-\hat{s}_{1})\hat{s}_{4}\hat{s}_{3})\\
 -(\frac{\sqrt{3}}{4}(\hat{s}_{1}^{2}+\hat{s}_{2}^{2})\hat{s}_{4}+\frac{\sqrt{3}}{2}\hat{s}_{1}\hat{s}_{2}\hat{s}_{4})\\
\end{bmatrix}\quad +\quad
\begin{bmatrix}
\frac{-3}{\sqrt{2}}\hat{s}_{1}^{2}\\
\-\sqrt{\frac{3}{2}}\hat{s}_{2}^{2}\\
-\sqrt{3}\hat{s}_{2}^{2}\\
\frac{-3}{4}\hat{s}_{1}\hat{s}_{2}\\
\frac{3}{2}i\hat{s}_{2}\hat{s}_{4}\\
-i\frac{\sqrt{3}}{2}(\hat{s}_{1}+\hat{s}_{2})\hat{s}_{4}\\
\frac{3}{2}\hat{s}_{1}\hat{s}_{2}\\
\frac{\sqrt{3}}{2}(\hat{s}_{3}^{2}-\hat{s}_{4}^{2})\\
\frac{\sqrt{3}}{2}(\hat{s}_{3}^{2}-\hat{s}_{4}^{2})\\
\frac{\sqrt{3}}{2}i(\hat{s}_{1}-\hat{s}_{2})\hat{s}_{4}\\
\frac{3}{2}\hat{s}_{1}\hat{s}_{4}\\
-\frac{3}{2}i\hat{s}_{2}\hat{s}_{3}\\
\frac{\sqrt{3}}{2}i(\hat{s}_{4}^{2}-\hat{s}_{3}^{2}-\hat{s}_{2}^{2}-(\hat{s}_{1}\hat{s}_{2}-2i\hat{s}_{4}\hat{s}_{3}))\\
\frac{\sqrt{3}}{2}i\hat{s}_{3}(\hat{s}_{2}-\hat{s}_{1})\\
\frac{\sqrt{3}}{2}i(\hat{s}_{4}^{2}-\hat{s}_{3}^{2}-\hat{s}_{2}^{2}+\hat{s}_{1}\hat{s}_{2}-2i\hat{s}_{4}\hat{s}_{3})\\
-\frac{\sqrt{3}}{2}i\hat{s}_{3}(\hat{s}_{2}+\hat{s}_{1})\\
\end{bmatrix}\quad+\quad
\begin{matrix}
\begin{bmatrix}
\sqrt{2}\hat{s}_{1}\\
\frac{2}{\sqrt{3}}\hat{s}_{1}\\
2\hat{s}_{2}\\
-\hat{s}_{3}\\
0\\
0\\
0\\
0\\
0\\
0\\
0\\
-\hat{s}_{4}\\
-\sqrt{3}i\hat{s}_{2}\\
\frac{\sqrt{3}}{2}\hat{s}_{4}\\
\sqrt{3}i\hat{s}_{2}\\
\sqrt{3}\hat{s}_{4}\\
\end{bmatrix}
\end{matrix}
\end{equation}
\end{widetext}
The $4\times4$ reduced density matrix of the $16\times1$ Stokes vector can be written as the following
\begin{equation}
\rho^{(2)}= \frac{1}{2}\sum_{N=1}^{16}\Omega_{N}S_{N}
\end{equation}.
This reduced density becomes important when we consider physical systems in which third harmonic generation, sum frequency generation, anti Stokes Raman scattering etc occurs. The $\rho$ matrix is calculated by substituting the first column of triple Stokes vector in Eq.(3.8) and by doing the summation over all existing elements of $4\times4$ analogue of Pauli matrices. Since the calculation of density matrix elements are tedious and time consuming we have used Mathematica to do complicated calculations. The derived $\rho$ (3.9) matrix has the form,
\begin{widetext}
\begin{equation}
\rho=\left(
\begin{array}{cccc}
E_{H}^{3}E_{H}^{*3} & E_{H}^{3}E_{V}^{* 3}& \sqrt{3}E_{H}^{3}E_{H}^{*}E_{V}^{* 2} & \sqrt{3}E_{H}^{3}E_{H}^{* 2}E_{V}^{*}\\
 E_{V}^{3}E_{H}^{*3}& E_{V}^{3}E_{V}^{*3} & \sqrt{3}E_{V}^{3}E_{H}^{*}E_{V}^{* 2}   &  \sqrt{3}E_{V}^{3}E_{H}^{* 2}E_{V}^{*} \\
 \sqrt{3}E_{H}E_{V}^{2}E_{H}^{* 3} &  \sqrt{3}E_{H}E_{V}^{2}E_{V}^{* 2} & 3 E_{H}E_{V}^{2}E_{H}^{*}E_{V}^{* 2}& 3E_{H}E_{V}^{2}E_{H}^{*2}E_{V}^{*} \\
 \sqrt{3}E_{H}^{2}E_{V}E_{H}^{* 3}&  \sqrt{3}E_{H}^{2}E_{V}E_{V}^{* 3} & 3E_{H}^{2}E_{V}E_{H}^{*}E_{V}^{* 2} & 3E_{H}^{2}E_{V}E_{H}^{*2}E_{V}^{*} \\
  \end{array}
\right)_{timeavg}
\end{equation}
\end{widetext}
As we can see the the off diagonal elements of the density matrix are conjugate of each other.\\

The $4\times16$ Mueller matrix of a nonlinear sample can be obtained using PIPO (Polarization-in Polarization-out)arrangement. The light first passes through PSG(Polarization State Generator), then through sample, then through PSA(Polarization State Analyzer). Light coming out of PSG  is represented using $16\times1$ Stokes vector. The PSG contains the optical elements such as polarizer, quarter wave plate and half wave plate. A combination of these optics gives freedom to move polarization states any where along the surface of Poincare sphere. The PSA(Polarization State Analyser) also contains the same optical components as in PSG. Sixteen polarization basis vectors are produced by PSG which is incident on the sample.This sixteen polarization basis vectors are chosen from the Poincare sphere. These points are lying symmetrically on the surface of the Poincare sphere. The emerging light from the sample is represented using $4\times1$ Stokes vector.The Stokes vector can be written in terms of coordinates of Poincare sphere (3.10). A point on the surface of the sphere is written as $s(\varepsilon,\theta)$ $\varepsilon$ is the latitude angle $\theta$ is the longitude angle of the sphere\cite{polarbook}.
 \begin{equation}
 \begin{bmatrix}
 s_1\\
 s_2\\
 s_3\\
 s_4\\
 \end{bmatrix}
 =
 \begin{bmatrix}
 1\\
 cos2\varepsilon cos2\theta\\
 cos2\varepsilon sin2\theta\\
 sin2\varepsilon
 \end{bmatrix}
 \end{equation}

Among the sixteen points on the surface of the sphere, six points are on the equator, which includes horizontal $s(0,0)$, vertical $s(0,\frac{\pi}{2})$ polarizations. The other points $s(0,\frac{\pi}{6})$,$s(0,\frac{\pi}{3})$,$s(0,-\frac{\pi}{3})$,$s(0,-\frac{\pi}{6})$ among six points are spaced at angle of $2\varepsilon=60^{\circ}$ from each other.These six points are all linearly polarized.
The other two points are North Pole (Right Circular Polarization,$s(\frac{\pi}{2},0)$) and South Pole (Left Circular Polarization,$s(\frac{-\pi}{2},0)$). The other eight points are elliptically polarized light. Four of them lies on the upper hemisphere equally spaced with $2\varepsilon$ equal to $45^{\circ}$ and $2\theta$ equally spaced at $45^{\circ}$. The points on the upper hemisphere are $s(\frac{\pi}{8},\frac{\pi}{8})$,$s(\frac{\pi}{8},\frac{3\pi}{8})$,
$s(\frac{\pi}{8},\frac{-\pi}{8})$,$s(\frac{\pi}{8},-\frac{3\pi}{8})$  Other four lies on the lower hemisphere at equally spaced points with $2\varepsilon$ equal to $-45^{\circ}$ and $2\theta$ equally spaced at $45^{\circ}$. Those points are $s(-\frac{\pi}{8},\frac{\pi}{8})$,$s(-\frac{\pi}{8},\frac{3\pi}{8})$.
$s(-\frac{\pi}{8},\frac{-\pi}{8})$,$s(-\frac{\pi}{8},-\frac{3\pi}{8})$\\. By substituting (3.10, 3.11) into (3.7), we obtain triple Stokes vector in terms of the coordinates of Poincare sphere. Using it we obtain the triple Stokes vector for the polarization basis vectors(3.12). Linear Stokes vectors for each of the sixteen polarization states chosen are given in Eq.(3.11). In order to determine a Mueller matrix of a given sample, light is incident on the sample in each of the sixteen polarization states. In each case, linear Stokes vector of the scattered light is measured by using four measurements. These 64 measurements give us elements of $4\times16$ Mueller matrix. Mueller matrix elements are related to components of susceptibility, which inturn are related to the third order nonlinear optical properties of the material.

\begin{equation}
 \begin{array} {|c||c|c|c|c|c|c|c|c|c|}
 \hline
  s(\varepsilon,\theta)   & s_1 & s_2 & s_3 & s_4 &  s(\varepsilon,\theta) & s_1 & s_2 & s_3 & s_4\\

     \hline \hline
 s(0,0)& 1 & 1 & 0 & 0 & s(\frac{\pi}{8}\frac{\pi}{8}) & 1  & \frac{1}{2} & \frac{1}{2} & \frac{1}{\sqrt{2}} \\

 s(0,\frac{\pi}{2}) & 1 &-1 & 0 & 0 & s(\frac{-\pi}{8},\frac{\pi}{8})& 1 & \frac{1}{2} & \frac{1}{2} & \frac{-1}{\sqrt{2}}\\

 s(0,\frac{\pi}{6})& 1  &\frac{1}{2} & \frac{\sqrt{3}}{2} & 0 & s(\frac{\pi}{8},\frac{3\pi}{8}) & 1 & \frac{-1}{2} & \frac{1}{2} & \frac{1}{\sqrt{2}}  \\

 s(0,\frac{\pi}{3})& 1 & \frac{-1}{2} & \frac{\sqrt{3}}{2} & 0   & s(\frac{-\pi}{8},\frac{3\pi}{8}) & 1&\frac{-1}{2} & \frac{1}{2} & \frac{1}{\sqrt{2}} \\

 s(0,-\frac{\pi}{3}) & 1 & \frac{-1}{2} & \frac{-\sqrt{3}}{2} & 0 & s(\frac{\pi}{8},\frac{-\pi}{8}) & 1  & \frac{1}{2} & \frac{-1}{2} &\frac{1}{\sqrt{2}}  \\

 s(0,-\frac{\pi}{6})& 1 & \frac{1}{2} & \frac{-\sqrt{3}}{2} & 0 & s(\frac{-\pi}{8},\frac{-\pi}{8}) &  1&\frac{1}{2} &  \frac{-1}{2} & \frac{-1}{\sqrt{2}}  \\

 s(\frac{-\pi}{4},0) & 1 & 0 & 0 &-1 & s(\frac{\pi}{8},\frac{-3\pi}{8}) & 1 & \frac{-1}{2} & \frac{-1}{2} & \frac{1}{\sqrt{2}}  \\

 s(\frac{\pi}{4},0) & 1 & 0 & 0 & -1 &  s(\frac{-\pi}{8},\frac{-3\pi}{8}) & 1 & \frac{-1}{2}& \frac{-1}{2} & \frac{-1}{\sqrt{2}}  \\

\hline
 \end{array}
 \end{equation}
 
 \begin{widetext}
 \begin{equation}
 \begin{array}{|c||c|c|c|c|c|c|c|c|c|c|c|c|c|c|c|c|}
   \hline
   S(\varepsilon,\theta)& S_{1} & S_{2} &  S_{3} & S_{4} & S_{5} & S_{6} & S_{7} & S_{8} & S_{9} & S_{10} & S_{11} & S_{12} & S_{13} & S_{14} & S_{15} & S_{16}\\
   \hline\hline

   S(0,0) & \frac{1}{\sqrt{2}} & \frac{1}{\sqrt{6}} & \frac{1}{\sqrt{3}} & 1 & 0 & 0 & 0 & 0 & 0 & 0 & 0 & 0 & 0 & 0 & 0 & 0 \\

   S(0,\frac{\pi}{2}) &  \frac{1}{\sqrt{2}}&  \frac{1}{\sqrt{6}} &  \frac{1}{\sqrt{3}} & -1 & 0 & 0 & 0 & 0 & 0 & 0 & 0 & 0 & 0 & 0 & 0 & 0 \\

   S(0,\frac{\pi}{6}) & \frac{1}{\sqrt{2}} & \frac{-11}{16\sqrt{6}} & \frac{5}{32\sqrt{3}} & \frac{13}{32} & \frac{3\sqrt{3}}{32} & \frac{3}{32} & \frac{9\sqrt{3}}{32} & \frac{9\sqrt{3}}{32} &\frac{3\sqrt{3}}{32}  &\frac{27}{32} & 0 & 0 & 0 & 0 & 0 & 0 \\

   S(0,\frac{\pi}{3})& \frac{1}{\sqrt{2}} & \frac{7}{16\sqrt{6}} & \frac{-13}{32\sqrt{3}} & \frac{-13}{32} & \frac{3\sqrt{3}}{32} & \frac{27}{32} &\frac{9\sqrt{3}}{32}  & \frac{3\sqrt{3}}{32} & \frac{9\sqrt{3}}{32} & \frac{-3}{32} & 0 & 0 & 0 & 0 & 0 & 0 \\

   S(0,-\frac{\pi}{3}) & \frac{1}{\sqrt{2}} & \frac{7}{16\sqrt{6}} & \frac{-13}{32\sqrt{3}} & \frac{-13}{32} & -\frac{3\sqrt{3}}{32} & \frac{-27}{32} & -\frac{9\sqrt{3}}{32} & \frac{3\sqrt{3}}{32} & \frac{9\sqrt{3}}{32} & \frac{-3}{32} & 0 & 0 & 0 & 0 & 0 & 0 \\

   S(0,-\frac{\pi}{6}) & \frac{1}{\sqrt{2}} & \frac{-11}{16\sqrt{6}} & \frac{5}{32\sqrt{3}} & \frac{13}{32} & -\frac{3\sqrt{3}}{32} & -\frac{3}{32} & -\frac{9\sqrt{3}}{32}  &  \frac{9\sqrt{3}}{32} & \frac{3\sqrt{3}}{32} & -\frac{27}{32} & 0 & 0 & 0 & 0 & 0 & 0 \\

   S(\frac{-\pi}{4},0) &  \frac{1}{\sqrt{2}} & \frac{-1}{2\sqrt{6}} & \frac{-1}{2\sqrt{3}} & 0 & 0 & 0 & 0 & 0 & \frac{-\sqrt{3}}{4} & 0 & \frac{-3}{4} & \frac{-1}{4} & 0 & \frac{-\sqrt{3}}{4} & 0 & \frac{\sqrt{3}}{4} \\

   S(\frac{\pi}{4},0) &  \frac{1}{\sqrt{2}} & \frac{-1}{2\sqrt{6}} & \frac{-1}{2\sqrt{3}} & 0 & 0 & 0 & 0 & 0 & \frac{-\sqrt{3}}{4} & 0 & \frac{3}{4} & \frac{1}{4} & 0 & \frac{\sqrt{3}}{4} & 0 & \frac{-\sqrt{3}}{4} \\

   S(\frac{\pi}{8},\frac{\pi}{8}) & \frac{1}{\sqrt{2}}&\frac{-11}{16\sqrt{6}} & \frac{5}{32\sqrt{3}} & \frac{13}{32} & \frac{-5}{32} & \frac{\sqrt{3}}{32} & \frac{9}{32} & \frac{-3\sqrt{3}}{32} & \frac{-\sqrt{3}}{32} & \frac{9\sqrt{3}}{32} & \frac{9}{16\sqrt{2}} & \frac{-1}{16\sqrt{2}} & \frac{-3\sqrt{3}}{8\sqrt{2}} & \frac{\sqrt{3}}{16\sqrt{2}} & \frac{\sqrt{3}}{8\sqrt{2}} & \frac{-9\sqrt{3}}{16\sqrt{2}} \\

   S(-\frac{\pi}{8},\frac{\pi}{8}) & \frac{1}{\sqrt{2}}&\frac{-11}{16\sqrt{6}} & \frac{5}{32\sqrt{3}} & \frac{13}{32} & \frac{-5}{32} & \frac{\sqrt{3}}{32} & \frac{9}{32} & \frac{-3\sqrt{3}}{32} & \frac{-\sqrt{3}}{32} & \frac{9\sqrt{3}}{32} & -\frac{9}{16\sqrt{2}} & \frac{1}{16\sqrt{2}} & \frac{3\sqrt{3}}{8\sqrt{2}} & \frac{-\sqrt{3}}{16\sqrt{2}} & \frac{-\sqrt{3}}{8\sqrt{2}} & \frac{-9\sqrt{3}}{16\sqrt{2}} \\

   S(\frac{\pi}{8},\frac{3\pi}{8})& \frac{1}{\sqrt{2}} & \frac{7}{16\sqrt{6}} & \frac{-13}{32\sqrt{3}} & \frac{-13}{32} & \frac{-5}{32} & \frac{9\sqrt{3}}{32} & \frac{9}{32}& \frac{-\sqrt{3}}{32} & \frac{-3\sqrt{3}}{32} & \frac{\sqrt{3}}{32} & \frac{9}{16\sqrt{2}} & \frac{-1}{16\sqrt{2}} & \frac{-\sqrt{3}}{8\sqrt{2}} & \frac{9\sqrt{3}}{16\sqrt{2}} & \frac{3\sqrt{3}}{8\sqrt{2}} & \frac{-\sqrt{3}}{16\sqrt{2}} \\

   S(-\frac{\pi}{8},\frac{3\pi}{8}) & \frac{1}{\sqrt{2}} & \frac{7}{16\sqrt{6}} & \frac{-13}{32\sqrt{3}} & \frac{-13}{32} & \frac{-5}{32} & \frac{9\sqrt{3}}{32} & \frac{9}{32}& \frac{-\sqrt{3}}{32} & \frac{-3\sqrt{3}}{32} & \frac{\sqrt{3}}{32} & \frac{-9}{16\sqrt{2}} & \frac{1}{16\sqrt{2}} & \frac{\sqrt{3}}{8\sqrt{2}} & \frac{-9\sqrt{3}}{16\sqrt{2}} & \frac{-3\sqrt{3}}{8\sqrt{2}} & \frac{\sqrt{3}}{16\sqrt{2}} \\

   S(\frac{\pi}{8},-\frac{\pi}{8}) & \frac{1}{\sqrt{2}}&\frac{-11}{16\sqrt{6}} & \frac{5}{32\sqrt{3}} & \frac{13}{32} & \frac{5}{32} & \frac{-\sqrt{3}}{32} & \frac{-9}{32} & \frac{-3\sqrt{3}}{32} & \frac{-\sqrt{3}}{32} & \frac{-9\sqrt{3}}{32} & \frac{9}{16\sqrt{2}} & \frac{-1}{16\sqrt{2}} & \frac{-3\sqrt{3}}{8\sqrt{2}} & \frac{\sqrt{3}}{16\sqrt{2}} & \frac{-\sqrt{3}}{8\sqrt{2}} & \frac{-9\sqrt{3}}{16\sqrt{2}} \\

   S(\frac{-\pi}{8},\frac{-\pi}{8}) & \frac{1}{\sqrt{2}}&\frac{-11}{16\sqrt{6}} & \frac{5}{32\sqrt{3}} & \frac{13}{32} & \frac{5}{32} & \frac{-\sqrt{3}}{32} & \frac{-9}{32} & \frac{-3\sqrt{3}}{32} & \frac{-\sqrt{3}}{32} & \frac{-9\sqrt{3}}{32} & \frac{-9}{16\sqrt{2}} & \frac{1}{16\sqrt{2}} & \frac{-3\sqrt{3}}{8\sqrt{2}} & \frac{-\sqrt{3}}{16\sqrt{2}} & \frac{-\sqrt{3}}{8\sqrt{2}} & \frac{9\sqrt{3}}{16\sqrt{2}} \\

   S(\frac{\pi}{8},\frac{-3\pi}{8})& \frac{1}{\sqrt{2}} & \frac{7}{16\sqrt{6}} & \frac{-13}{32\sqrt{3}} & \frac{-13}{32} & \frac{5}{32} & \frac{-9\sqrt{3}}{32} & \frac{-9}{32}& \frac{-\sqrt{3}}{32} & \frac{-3\sqrt{3}}{32} & \frac{-\sqrt{3}}{32} & \frac{9}{16\sqrt{2}} & \frac{-1}{16\sqrt{2}} & \frac{\sqrt{3}}{8\sqrt{2}} & \frac{9\sqrt{3}}{16\sqrt{2}} & \frac{-3\sqrt{3}}{8\sqrt{2}} & \frac{-\sqrt{3}}{16\sqrt{2}} \\

    S(\frac{-\pi}{8},\frac{-3\pi}{8})& \frac{1}{\sqrt{2}} & \frac{7}{16\sqrt{6}} & \frac{13}{32\sqrt{3}} & \frac{-13}{32} & \frac{5}{32} & \frac{-9\sqrt{3}}{32} & \frac{-9}{32}& \frac{-\sqrt{3}}{32} & \frac{-3\sqrt{3}}{32} & \frac{-\sqrt{3}}{32} & \frac{-9}{16\sqrt{2}} & \frac{1}{16\sqrt{2}} & \frac{\sqrt{3}}{8\sqrt{2}} & \frac{-9\sqrt{3}}{16\sqrt{2}} & \frac{3\sqrt{3}}{8\sqrt{2}} & \frac{\sqrt{3}}{16\sqrt{2}} \\

   \hline
 \end{array}
 \end{equation}
 \end{widetext}

The  linear Stokes vector for these sixteen polarization basis vectors are given(4.3). This basis vectors can be used to analyze sample in PIPO arrangement. The Mueller matrix elements can be related to susceptibility components\cite{Masood2015}. The susceptibility tensor elements gives information about the property of that material.
 
\section{Conclusion}

     Stokes-Mueller formalism recently has become an important tool in characterizing nonlinear optical properties of materials. Nonlinear-Stokes Mueller polarimetry has the advantage of measuring all the susceptibility components of the sample in the given configuration at the same instant. In this report we have developed a general Stokes formalism for a three photon process which is not discussed in the manner we did in the literature so far. The power of this triple Stokes formalism can be gauged from the fact that considering three photon process leads to the description and understanding third order nonlinear optical phenomena such as third harmonic generation, saturable absorption, coherent anti Stokes Raman scattering, etc. Triple Stokes vector is derived analytically using quantized form of light as a $16\times1$ vector. A total sixteen independent initial polarization states of light are obtained using points on Poincare sphere. PIPO experiments on scattering involving three photon processes can be used to experimentally determine the $4\times16$ Mueller matrix of a given sample.

\end{document}